\RequirePackage{fix-cm}
\documentclass[twocolumn]{svjour3}          
\smartqed  
\usepackage{graphicx}
\usepackage{hyperref}
\begin{document}

\title{Impact of lepton $p_{T}$ threshold on the charge asymmetry predictions for the inclusive $W$ boson production in $pp$ collisions at 13 TeV
}


\author{Kadir Ocalan}


\institute{K. Ocalan \at
              Necmettin Erbakan University, Faculty of Aviation and Space Sciences, Konya, Turkey  \\
              ORCID ID: http://orcid.org/0000-0002-8419-1400 \\
              \email{kadir.ocalan@erbakan.edu.tr}           
}

\date{Received: date / Accepted: date}

\maketitle

\begin{abstract}
This paper presents the impact of lepton transverse momentum $p^{l}_{T}$ threshold on the $W$ boson charge asymmetry predictions in perturbative QCD for the inclusive $W^{\pm}+X \rightarrow l^{\pm} \nu +X$ production in proton-proton ($pp$) collisions. The predictions are obtained with various low-$p^{l}_{T}$ thresholds $p^{l}_{T} >$ 20, 25, 30, and 40 GeV in a fiducial region encompassing both central and forward detector acceptances in terms of the lepton pseudorapidty 0 $\leq \eta_{l} \leq$ 4.5. The predicted distributions for the lepton charge asymmetry, which is defined by means of the $\eta_{l}$ ($A_{\eta_{l}}$), at next-to-next-to-leading order (NNLO) accuracy are compared with the CMS and LHCb data at 8 TeV center-of-mass collision energy. The 8 TeV predictions are observed to reproduce the data fairly well within the quoted uncertainties. The predictions from the CT14 parton distribution function (PDF) model provide slightly better agreement with the data over the other PDF sets that are being tested. The 13 TeV predictions by using various $p^{l}_{T}$ thresholds are reported for the $A_{\eta_{l}}$ and also the charge asymmetries that are defined in terms of the differential cross sections in bins of the $W$ boson rapidity $y_{W}$ ($A_{y_{W}}$) and transverse momentum $p^{W}_{T}$ ($A_{p^{W}_{T}}$). The NNLO predictions for the $A_{\eta_{l}}$, $A_{y_{W}}$, and $A_{p^{W}_{T}}$ distributions are assessed to be in close correlation with the $p^{l}_{T}$ value. The $A_{\eta_{l}}$ and $A_{y_{W}}$ distributions are particularly shown to be more correlated at a higher $p^{l}_{T}$ threshold. The $A_{p^{W}_{T}}$ distributions are also reported from the merged predictions with improved accuracy by the inclusion of next-to-next-to-next-to-leading logarithm (N$^{3}$LL) corrections, i.e., at NNLO+N$^{3}$LL. The predicted distributions from various $p^{l}_{T}$ thresholds represent also a finer probe in terms of the capability to provide more constraints on the ratio of $u$ and $d$ quark distribution functions in the parton momentum fraction range $10^{-4} < x < 1$.
  
\end{abstract}

\section{Introduction}
\label{intro}
Weak vector boson ($W$ and $Z$ boson) production plays a crucial role at hadron colliders including the present proton--proton ($pp$) collider at CERN, the Large Hadron Collider (LHC). Production of $W$ and $Z$ bosons in $pp$ collisions at the LHC enables several precision tests of the quantum chromodynamic (QCD) and electroweak (EW) sectors of the Standard Model (SM). Their precise measurements provide substantial inputs for constraining parton distribution functions (PDFs) in the proton and improved background modeling for several rarer SM processes such as top quark and Higgs boson productions and beyond the SM searches such as supersymmetry and dark matter. Their measurements in leptonic decay modes are very advantageous as they are produced in abundance with clean experimental signatures and constitute a major experimental benchmark to calibrate detector response for lepton, jet, and missing transverse energy reconstructions. Their productions through leptonic decays are not only important for experimental aspects, but they are also essential to test Monte Carlo based event generators and fixed-order calculations for the advancement of the field of theoretical predictions.  

In particular, $W$ boson production is experimentally characterized by one isolated lepton with high transverse momentum $p_{T}$ and large missing transverse energy due to neutrino in its leptonic decay mode $pp \rightarrow W^{\pm} \rightarrow l^{\pm}\nu$, where $l$ is either a muon $\mu$ or an electron $e$. Dominant mechanism for $W$ boson production at the LHC proceeds via annihilation of a valence quark from one of colliding protons with a sea antiquark from other proton as $u\bar{d}\rightarrow W^{+}$ and $d\bar{u}\rightarrow W^{-}$. The excess of two valence $u$ quarks over one valence $d$ quark in the proton requires $W^{+}$ bosons to be produced more often than $W^{-}$ bosons. This production asymmetry between $W^{+}$ and $W^{-}$ bosons is referred to as $W$ boson charge asymmetry and is usually defined with cross sections $\sigma(W^{+})$ and $\sigma(W^{-})$ differential in $W$ boson rapidity $y_{W}$ as     

\begin{equation}
\label{eqn:1}   
A_{y_{W}}=\frac{d\sigma(W^{+}\rightarrow l^{+}\nu)/dy_{W}-d\sigma(W^{-}\rightarrow l^{-}\bar{\nu})/dy_{W}}{d\sigma(W^{+}\rightarrow l^{+}\nu)/dy_{W}+d\sigma(W^{-}\rightarrow l^{-}\bar{\nu})/dy_{W}}.
\end{equation}
The $W$ boson charge asymmetry $A_{y_{W}}$ provides a direct probe of the relative $u$ and $d$ quark distributions as functions of the initial-state parton momentum fractions ($x$ values) since $y_{W}$ is strongly correlated with the $x$ values, which can generally be expressed as $x_{1, 2}=(M_{W}/\sqrt{s})e^{\pm y}$ with $M_{W}$ is the $W$ boson mass and $\sqrt{s}$ is the center-of-mass energy. However, there is an experimental limitation regarding the $A_{y_{W}}$ since the $p_{T}$ and $y_{W}$ of the $W$ boson cannot be directly reconstructed due to the unknown longitudinal momentum of the decay neutrino. Despite this limitation, the same information can still be accessed by measuring the charge asymmetry from the decay lepton. The charge asymmetry can readily be measured as a function of the decay lepton pseudorapidity $\eta_{l}$, which is indeed correlated with the $y_{W}$, in the analogous form of

\begin{equation}
\label{eqn:2}   
A_{\eta_{l}}=\frac{d\sigma(W^{+}\rightarrow l^{+}\nu)/d\eta_{l}-d\sigma(W^{-}\rightarrow l^{-}\bar{\nu})/d\eta_{l}}{d\sigma(W^{+}\rightarrow l^{+}\nu)/d\eta_{l}+d\sigma(W^{-}\rightarrow l^{-}\bar{\nu})/d\eta_{l}}.
\end{equation}         
The lepton charge asymmetry $A_{\eta_{l}}$ corresponds to the convolution of the original $A_{y_{W}}$ variable and the $V-A$ (vector--axial vector) asymmetry of the $W$ boson, which implies its anisotropic decay into the lepton and the neutrino. In a similar way, the $A_{\eta_{l}}$ variable can provide substantial constraints on the ratio of $u$ and $d$ quark distribution functions in the proton as a function of $x$ values of the partons. This variable can also be benefited for discriminating among various PDF models that predict different shapes of valence and sea quark distributions. 

The $W$ boson production asymmetries were measured before mostly in terms of the $A_{\eta_{l}}$ variable in p$\rm{\bar{p}}$ collisions by the CDF and D0 Collaborations at the Tevatron~\cite{Abe:1998rv,Abazov:2007pm,Abazov:2008qv,Aaltonen:2009ta,Abazov:2013rja,D0:2014kma}. The asymmetries were measured at the LHC by using the $A_{\eta_{l}}$ variable in the central lepton pseudorapidity region $|\eta_{l}|\leq$ 2.5 by the ATLAS and CMS Collaborations at different center-of-mass energies up to 8 TeV~\cite{Aad:2011dm,Chatrchyan:2012xt,Chatrchyan:2013mza,Khachatryan:2016pev,Aaboud:2016btc,Aad:2019bdc,Aaboud:2018nic,Aad:2019rou}. The $A_{\eta_{l}}$ variable was also measured by the LHCb Collaboration at the LHC up to 8 TeV~\cite{Aaij:2012vn,Aaij:2014wba,Aaij:2015zlq,Aaij:2016qqz} in the forward region 2.0 $\leq \eta_{l} \leq$ 4.5 extending beyond the ATLAS and CMS detector coverage. The entire LHC measurements have probed the inclusive $W$ boson cross sections along with the $A_{\eta_{l}}$ variable in the range $10^{-4} < x < 1$ which are clearly important to provide valuable inputs on determining accurate PDFs also at very small and large $x$ values. In all these measurements complementing in the $\mu$ and $e$ decay channels in terms of the wide $\eta_{l}$ region probed, data were compared with various theoretical predictions including fixed-order perturbative QCD calculations at next-to-leading order (NLO) and next-to-NLO (NNLO) accuracies, convolved with different PDF models.  

The $W$ boson charge asymmetries are being determined for kinematic phase spaces specified by the decay lepton transverse momentum $p^{l}_{T}$ threshold. The $p^{l}_{T}$ is correlated with the $W$ boson transverse momentum $p^{W}_{T}$, and hence, impacts in both measurements and predictions of the charge asymmetry. In the measurements, the $p^{l}_{T}$ threshold value is chosen to match with available detector triggering conditions and to have an efficient event reconstruction for a pure signal data sample with sufficiently high statistics. Thereby, the measurements are subject to use a $p^{l}_{T}$ threshold value which depends on event triggering and reconstruction requirements. Nevertheless in the theoretical calculations, various $p^{l}_{T}$ thresholds (including also the ones used in the measurements) can be used alternatively to test the impact on the charge asymmetry. Furthermore, theoretical predictions can be repeated with increasing thresholds in the low-$p^{l}_{T}$ region to select only a subset of phase space where the $\eta_{l}$ gets closer to the $y_{W}$. This also facilitates to test charge asymmetry predictions in a more constrained phase space in different ranges of the $\eta_{l}$ and $y_{W}$ allowing also a finer probe of the dependence to $x$ values. 

In this work we present the predicted charge asymmetries corresponding to the $W^{\pm}$ boson production processes $pp \rightarrow W^{+}+X \rightarrow l^{+}\nu+X$ and $pp \rightarrow W^{-}+X \rightarrow l^{-}\bar{\nu}+X$. The predictions are obtained in the fiducial phase space encompassing both central and forward regions 0 $\leq \eta_{l} \leq$ 4.5 at both 8 TeV and 13 TeV. The predictions at NNLO accuracy as a function of the $\eta_{l}$ from various PDF models are compared with the CMS and LHCb $pp$ collision data at 8 TeV. The predictions are further obtained as functions of the $\eta_{l}$ and $y_{W}$ at NNLO accuracy as well as in bins of the $p^{W}_{T}$ through resummation at next-to-next-to-next-to-leading logarithm (N$^{3}$LL) which is matched to NNLO, i.e., NNLO+N$^{3}$LL accuracy at 13 TeV. Various thresholds in low-$p^{l}_{T}$ region, $p^{l}_{T} >$ 20, 25, 30, and 40 GeV, are used to enable testing the potential impact on the charge asymmetry. Specifically the 13 TeV predictions are reported by aiming to assess the correlations among the increasing low-$p^{l}_{T}$ thresholds and the charge asymmetry distributions for the $A_{\eta_{l}}$, $A_{y_{W}}$, and also in bins of the $p^{W}_{T}$.

\section{Methodology}
\label{meth}
\subsection{Computational setup}
\label{comp}
The charge asymmetry calculations that are based on differential cross sections are performed by using the MATRIX framework~\cite{Grazzini:2017mhc,Catani:2009sm} which is interfaced with the RadISH program~\cite{Bizon:2017rah,Monni:2016ktx}, together by the computational framework MATRIX+RadISH (v1.0.1)~\cite{Kallweit:2020gva}. The fixed-order calculations of differential cross sections at NNLO in the QCD perturbation theory is achieved with the MATRIX framework which implements the transverse momentum \emph{$q_{T}$}-subtraction method~\cite{Catani:2007vq,Catani:2012qa}. In the \emph{$q_{T}$}-subtraction approach, infrared divergences of the real radiation contributions are extracted by using the infrared subtraction terms in the perturbative expansion. These divergences are regulated by employing a fixed cut-off value $r_{cut}=$ 0.0015 (0.15\%) for the slicing parameter $r$, where it is defined as $r=p_{T}/m$ in terms of the $p_{T}$ distribution and invariant mass $m$ for a system of colorless particles. The resummation of large logarithmic contributions, which is needed for accurate prediction of differential cross sections as a function of the $p^{W}_{T}$, is achieved with the formalism of the RadISH program. The RadISH code enables high-accuracy resummation for the $p^{W}_{T}$ distribution through N$^{3}$LL which is matched to the NNLO QCD calculations by the MATRIX. In addition, the OpenLoops tool~\cite{Cascioli:2011va,Denner:2016kdg} is utilized through an automated interface to acquire all the spin- and color-correlated tree-level and one-loop scattering amplitudes in the computations. In the setup, the Fermi constant $G_{F}$ input scheme is used where leptons (both $\mu$ and $e$) and light quarks are treated massless. The default MATRIX setup is used for the SM input parameters relevant to the inclusive $W$ boson process that are all based on the following $W$ boson mass and $G_{F}$ values of     
\begin{equation}
\label{eqn:3}   
M_{W}=80.385 \hspace{0.1cm} \rm{GeV}, \hspace{0.1cm} \it{G_{F}}=\rm{1.16639} \times \rm{10^{-5}} \hspace{0.1cm} \rm{GeV}^{-2}.
\end{equation}          
To this end, the QCD calculations of differential cross sections for the charge asymmetry predictions require inclusion of knowledge of PDFs. The evaluation of PDFs from data files is carried out by exploiting the LHAPDF (v6.2.0) framework~\cite{Buckley:2014ana} in the computations. Various PDF sets are used in the calculations, where all are based on a constant strong coupling $\alpha_{s}=$ 0.118. Particularly the NNLO PDF sets MMHT2014~\cite{Harland-Lang:2014zoa}, CT14~\cite{Dulat:2015mca}, NNPDF3.1~\cite{Ball:2014uwa}, and PDF4LHC15~\cite{Butterworth:2015oua} are used in the calculations.        

\subsection{Fiducial requirements}
\label{fid}
The calculations for both the differential cross section and charge asymmetry predictions are performed in a realistic fiducial phase space of the $W$ boson and its decay lepton. The fiducial phase space is defined to be in line with the reference CMS~\cite{Khachatryan:2016pev} and LHCb~\cite{Aaij:2015zlq} 8 TeV measurements. The leptons (either $\mu$ or $e$) are required to have transverse momentum $p^{l}_{T}>$ 25 GeV ($p^{l}_{T}>$ 20 GeV) and to lie in the $\eta_{l}$ region $0 \leq \eta_{l} \leq 2.4$ ($2.0 \leq \eta_{l} \leq 4.5$) for the validation of the predictions with the reference CMS (LHCb) results at 8 TeV. The leptons are required to have transverse momentum $p^{l}_{T}>$ 20 GeV and to lie in the $\eta_{l}$ region encompassing both central and forward acceptances $0 \leq \eta_{l} \leq 4.5$ at 13 TeV. In addition, the requirements $p^{l}_{T}>$ 25, 30, and 40 GeV are all used to assess the correlations of these increasing thresholds with the predicted charge asymmetries in the entire acceptance region $0 \leq \eta_{l} \leq 4.5$ concerning the 13 TeV predictions. Leptons are treated massless in the computational setup, thereby the predictions of differential cross sections in the $\mu$ channel are the same as in the $e$ channel. No requirements are strictly imposed for the $W$ boson transverse mass and the missing transverse energy due to the neutrino while these requirements can make more sense depending on experimental measurement. Moreover, no explicit requirement is applied for the final-state hadronic jet(s) in terms of jet definition criteria and selection cuts.           

\subsection{Theoretical uncertainties}
\label{theo}
Theoretical calculations of the cross sections in perturbative QCD expansions in the $\alpha_{s}$ depend on the choices for the renormalization $\mu_{R}$ and factorization $\mu_{F}$ scales. In this paper, the central values for the $\mu_{R}$ and $\mu_{F}$ scales are fixed to the $W$ boson mass $\mu_{R}$ = $\mu_{F}$ = $M_{W}$ = 80.385 GeV. Similarly the central value for the resummation scale $x_{Q}$ is set to the $W$ boson mass $x_{Q}$ = $M_{W}$ = 80.385 GeV when resummation of large logarithmic corrections is also considered in the calculations. Theoretical uncertainties due to the choices for the central scale values or shortly scale uncertainties correspond to missing higher-order corrections in the perturbative (and resummed) calculations. Scale uncertainties are estimated by varying independently the $\mu_{R}$ and $\mu_{F}$ by a factor of 2 up and down around their central values. The seven-point variation method is employed, that is all possible combinations in the variations are considered while imposing the constraint $0.5 \leq \mu_{R}/ \mu_{F} \leq 2.0$. On the other hand, the nine-point variation method is used when perturbative calculations include also matching to resummation, that is the envelope of the seven-point variation while keeping $x_{Q}$ at its central value and the two-point variation of $x_{Q}$ around its central value by a factor of 2 in either direction for the central values of the $\mu_{R}$ and $\mu_{F}$ scales. The PDF uncertainties due to different parametrization of the PDF models are estimated by following the prescription of the PDF4LHC working group~\cite{Butterworth:2015oua,Buckley:2014ana}. The $\alpha_{s}$ uncertainty is also estimated by varying the $\alpha_{s}$ value by $\pm0.001$ around 0.118. Thereafter, total theoretical uncertainties of the predictions are obtained by summing quadratically scale, PDF, and $\alpha_{s}$ uncertainties. Total theoretical uncertainties in the predicted distributions are presented symmetrically by using the larger values from estimated up and down uncertainties in a conservative consideration.

\section{Comparisons with the 8 TeV data }
\label{8tevcomp}
\begin{figure}
\includegraphics[width=9.00cm]{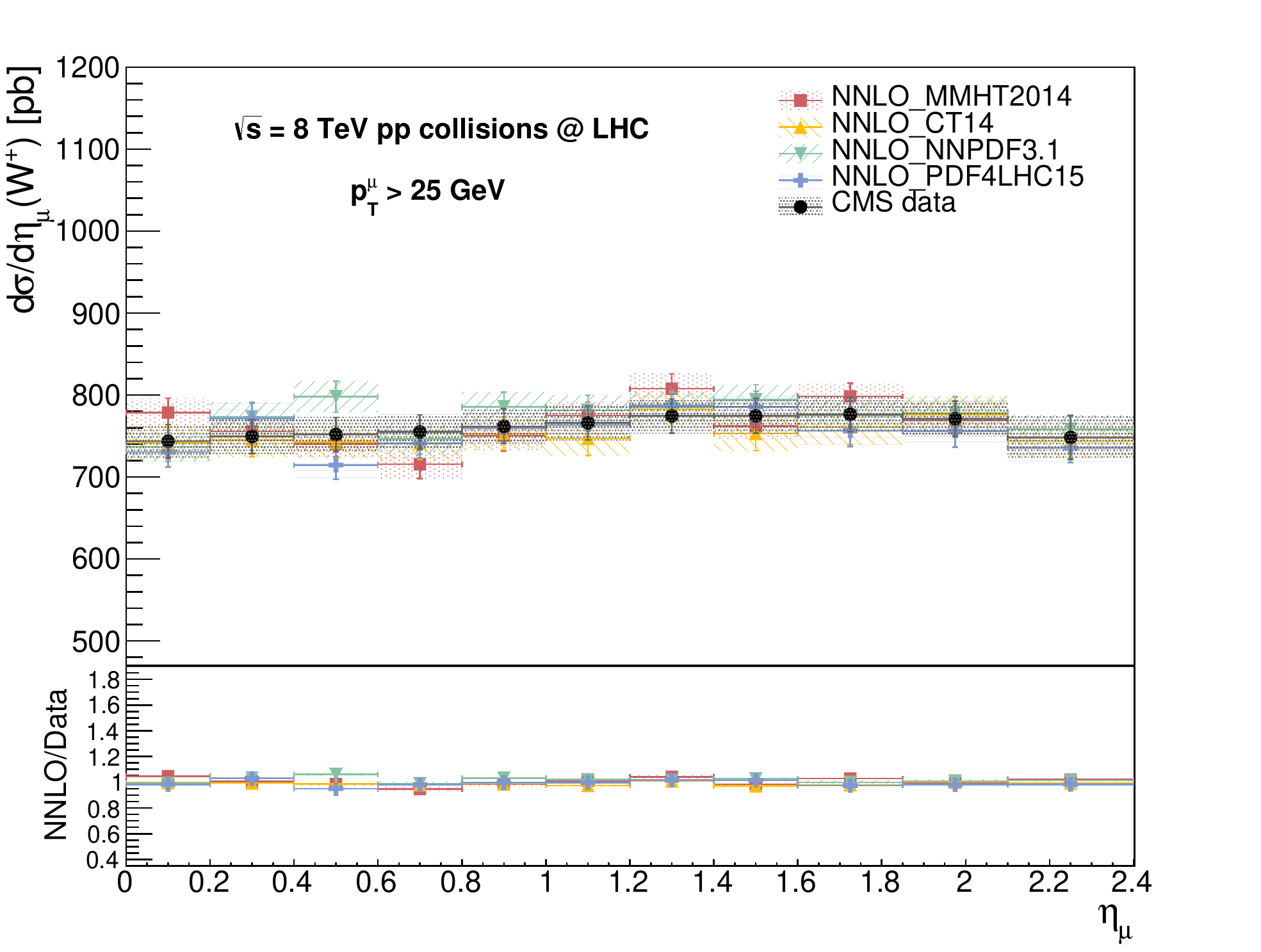}
\includegraphics[width=9.00cm]{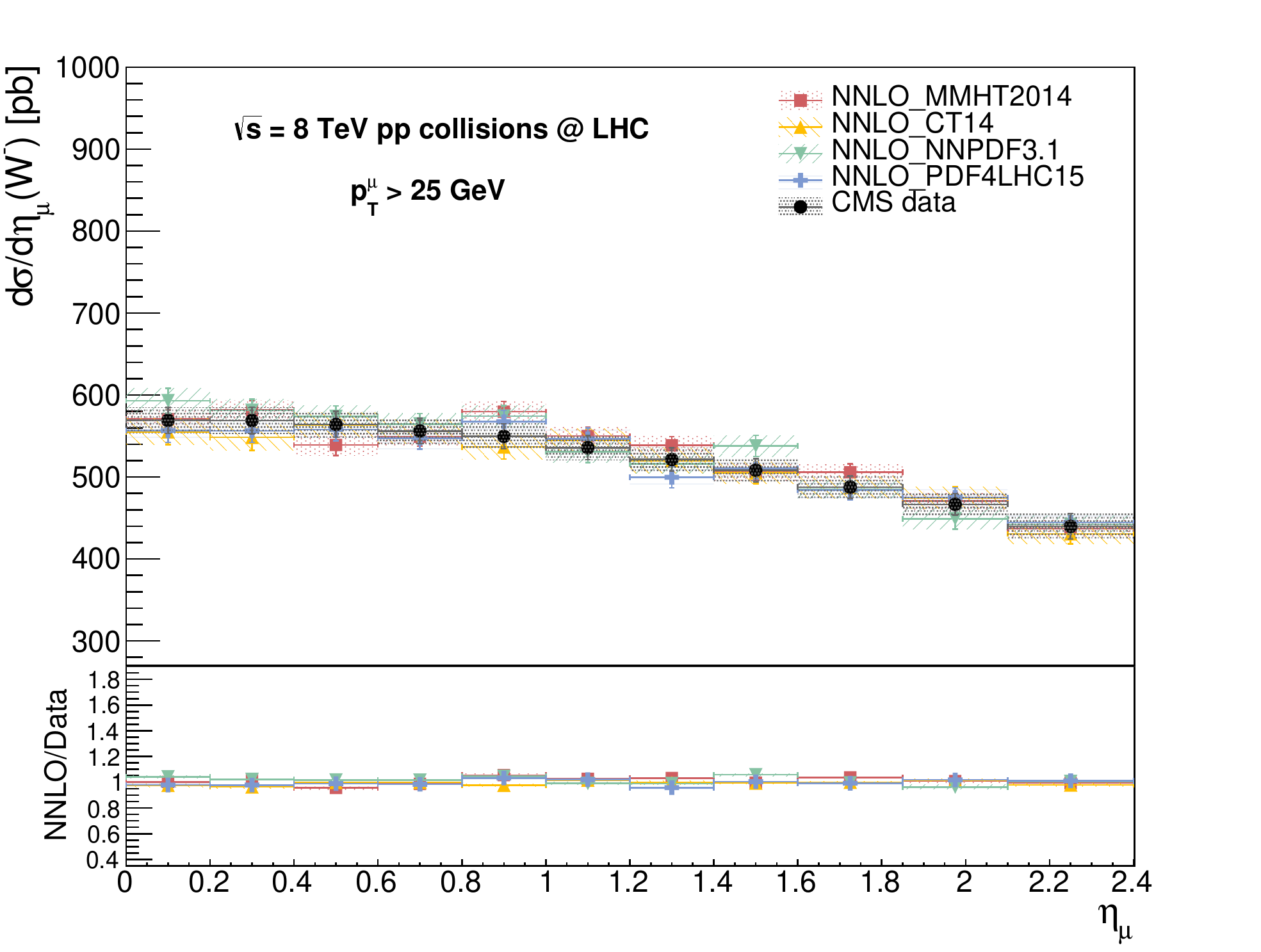}
\caption{The predicted differential cross section distributions for the $W^{+}$ (top) and $W^{-}$ (bottom) processes as a function of the $\eta_{\mu}$ and their comparisons with the CMS data at 8 TeV. The predictions are obtained at NNLO accuracy by using MMHT2014, CT14, NNPDF3.1, and PDF4LHC15 PDF sets. The predictions include total theoretical uncertainties from quadratic sum of scale, PDF, and $\alpha_s$ uncertainties, while the data include total experimental uncertainty. In the lower panels, the ratios of the predictions to the data for the differential cross section distributions are also displayed.}
\label{fig:1}      
\end{figure}

\begin{figure}
\includegraphics[width=9.00cm]{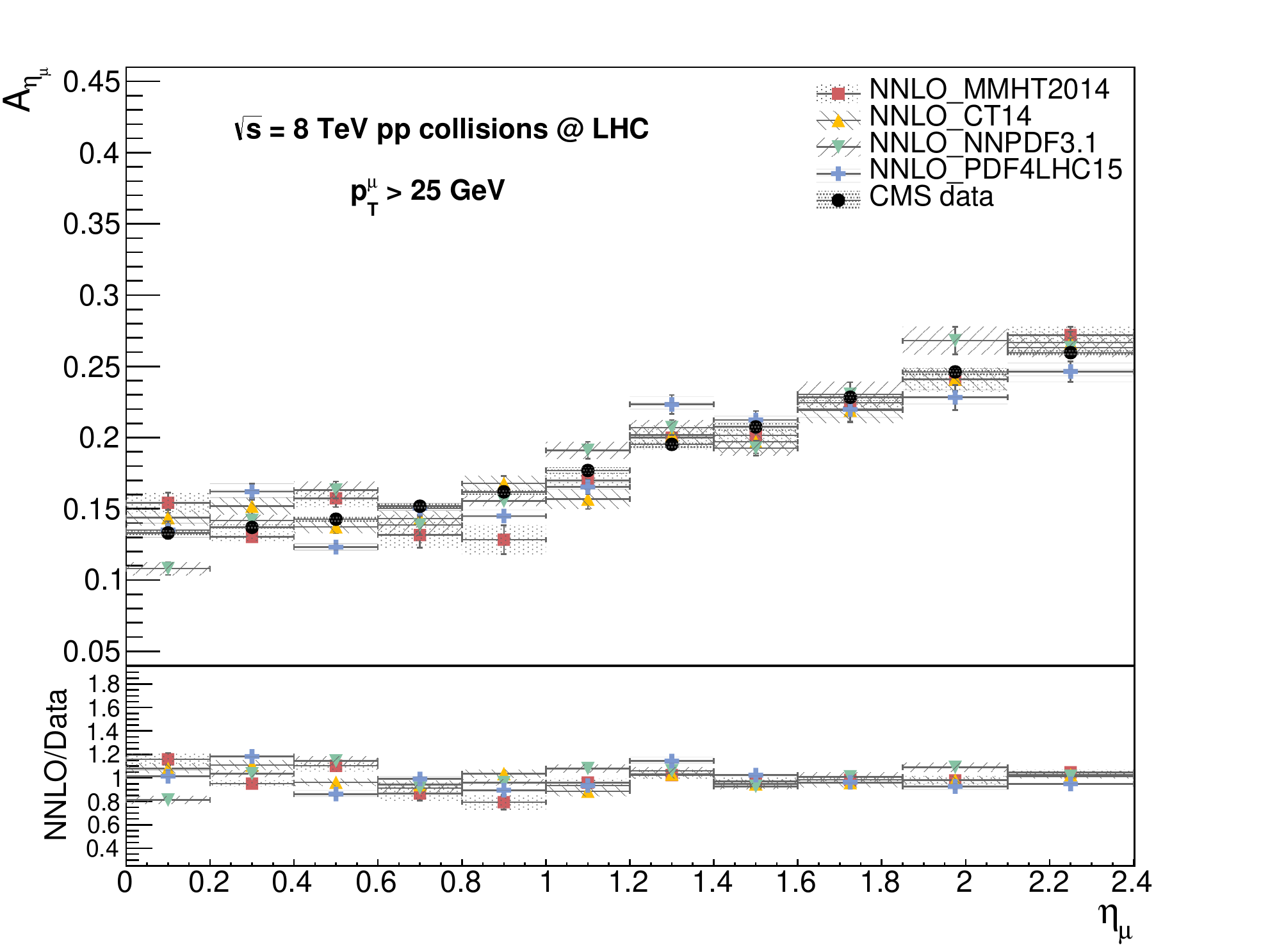}
\caption{The NNLO predictions for the muon charge asymmetry $A_{\eta_{\mu}}$ variable from MMHT2014, CT14, NNPDF3.1, and PDF4LHC15 PDF sets as a function of the $\eta_{\mu}$. The predictions are compared with the CMS data in the central region $0 \leq \eta_{\mu} \leq 2.4$ at 8 TeV. The predictions include total theoretical uncertainties from quadratic sum of scale, PDF, and $\alpha_s$ uncertainties, while the data include total experimental uncertainty. In the lower panel, the ratios of the predictions to the data for the $A_{\eta_{\mu}}$ are also displayed.}
\label{fig:2}       
\end{figure}

The predictions are compared with the 8 TeV data from the reference CMS measurement~\cite{Khachatryan:2016pev} for the differential cross section and $A_{\eta_{l}}$ distributions. The NNLO predictions are obtained in the $\mu$ decay mode with the fiducial requirement $p^{\mu}_{T}>$ 25 GeV in the central region $0 \leq \eta_{\mu} \leq 2.4$. The $\eta_{\mu}$ bin ranges are used from the CMS measurement as (0.00, 0.20), (0.20, 0.40), (0.40, 0.60), (0.60, 0.80), (0.80, 1.00), (1.00, 1.20), (1.20, 1.40), (1.40, 1.60), (1.60, 1.85), (1.85, 2.10), and (2.10, 2.40) to enable direct comparisons. Total theoretical uncertainties are included from quadratic sum of scale, PDF, and $\alpha_{s}$ uncertainties for the predicted distributions. Total experimental uncertainties are included by summing statistical, systematic, and luminosity uncertainties in quadrature for the measured differential cross sections, while summing statistical and systematic uncertainties in quadrature for the measured asymmetry. The NNLO predictions from the PDF sets MMHT2014, CT14, NNPDF3.1, and PDF4LHC15 are compared with the CMS data distributions. The predicted differential cross section distributions for the $W^{+}$ and $W^{-}$ processes are compared with the data in Fig.~\ref{fig:1}. The predictions using different PDF sets are observed to be in good agreement with each other and the data within uncertainties. The prediction using CT14 shows better agreement with the data, where the predictions using PDF sets other than CT14 slightly deviate from the data in only a small number of bins up to a few percent. The predicted $A_{\eta_{\mu}}$ distributions are compared with the data as shown in Fig.~\ref{fig:2}. Apart from a few exceptions, the predictions describe the CMS data consistently within uncertainties throughout the entire $\eta_{\mu}$ ranges. The predicted $A_{\eta_{\mu}}$ distribution from CT14 describes the data slightly better over the predictions using the other PDF sets. It can also be observed that sensitivity to discriminate among various PDF sets is enhanced in the $A_{\eta_{\mu}}$ variable in comparison to the differential cross sections. Moreover, the predicted results from various PDF sets for both the differential cross section and $A_{\eta_{\mu}}$ distributions are observed to be in agreement with the corresponding NNLO predictions by the FEWZ program~\cite{Li:2012wna} in the CMS paper. Difference between the MATRIX predictions and the FEWZ ones are generally up to $\sim$1--2\% within the quoted theoretical uncertainties.   

\begin{figure}
\includegraphics[width=9.00cm]{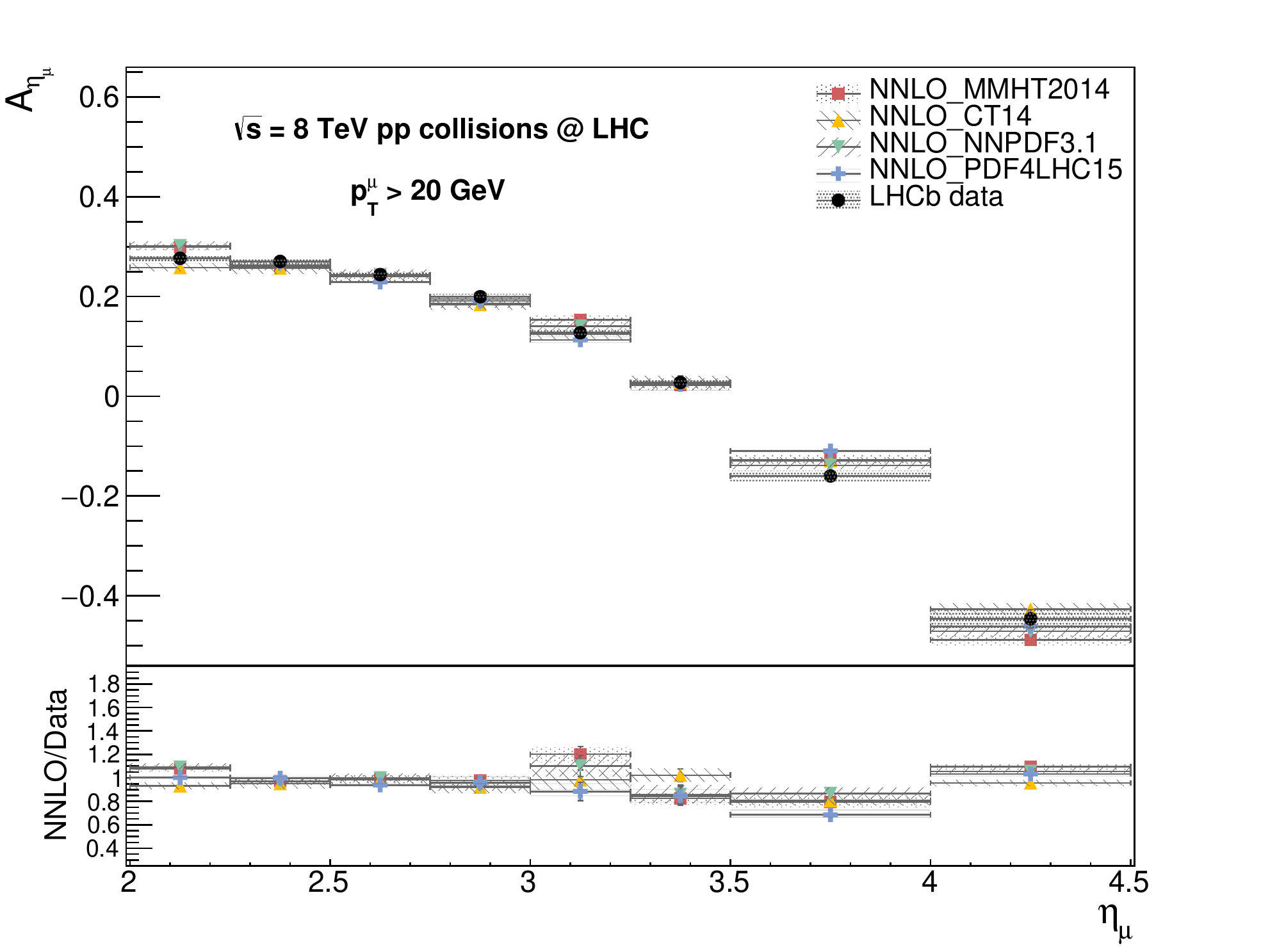}
\caption{The NNLO predictions for the $A_{\eta_{\mu}}$ variable from MMHT2014, CT14, NNPDF3.1, and PDF4LHC15 PDF sets as a function of the $\eta_{\mu}$. The predictions are compared with the LHCb data in the forward region $2.0 \leq \eta_{\mu} \leq 4.5$ at 8 TeV. The predictions include total theoretical uncertainties, while the data include total experimental uncertainty. In the lower panel, the ratios of the predictions to the data for the $A_{\eta_{\mu}}$ are also displayed.}
\label{fig:3}      
\end{figure}

The NNLO predictions are also compared with the 8 TeV data from the reference LHCb measurement~\cite{Aaij:2015zlq} which was performed in the $\mu$ decay mode in the forward acceptance region. Fiducial phase space requirement of $p^{\mu}_{T}>$ 20 GeV in the forward region $2.0 \leq \eta_{\mu} \leq 4.5$ is imposed to compare with the LHCb data for the $A_{\eta_{\mu}}$ variable. Bin edges for the $\eta_{\mu}$ are used identically from the LCHb measurement as (2.00, 2.25), (2.25, 2.50), (2.50, 2.75), (2.75, 3.00), (3.00, 3.25), (3.25, 3.50), (3.50, 4.00), and (4.00, 4.50). Total theoretical and experimental uncertainties are included to the central results for the predictions and data, respectively. Comparisons of the predicted $A_{\eta_{\mu}}$ distributions from various PDF sets with the data are shown in Fig.~\ref{fig:3}. The predictions are generally in good agreement with the data within uncertainties throughout the entire $\eta_{\mu}$ ranges. The prediction using CT14 tends to be slightly more consistent with the data over the results obtained using other PDF sets. The predicted results from all the PDF sets show no significant deviation from the FEWZ NNLO predictions that are presented in the LHCb measurement.       

To conclude here, the NNLO calculations are validated with the data for the predicted distributions of the differential cross sections and $A_{\eta_{l}}$ variable in the $\mu$ decay mode at 8 TeV. The predictions exhibit no significant deviations from the CMS and LHCb data within the quoted uncertainties in both central and forward regions $0 \leq \eta_{l} \leq 2.4$ and $2.0 \leq \eta_{l} \leq 4.5$. Predicted results obtained by using CT14 PDF set reproduce data more consistently among several PDF sets that are being tested. The predictions are also observed to be in agreement with the FEWZ NNLO results reported in the CMS and LHCb measurements. These 8 TeV comparisons encourage the extension of the NNLO calculations by the MATRIX+RadISH framework to 13 TeV, the current center-of-mass energy of the LHC, where the impact of several $p^{l}_{T}$ thresholds to the $W$ boson charge asymmetry variables can be assessed further. Validation of the NNLO calculations in the $e$ decay mode for the forward region $2.0 \leq \eta_{e} \leq 4.25$ by using the 8 TeV LHCb data was reported before in Ref.~\cite{Ocalan2021}.  

\section{The $A_{\eta_{l}}$ and $A_{y_{W}}$ predictions at 13 TeV}
\label{13tevpred}
\begin{figure}
\includegraphics[width=8.8cm]{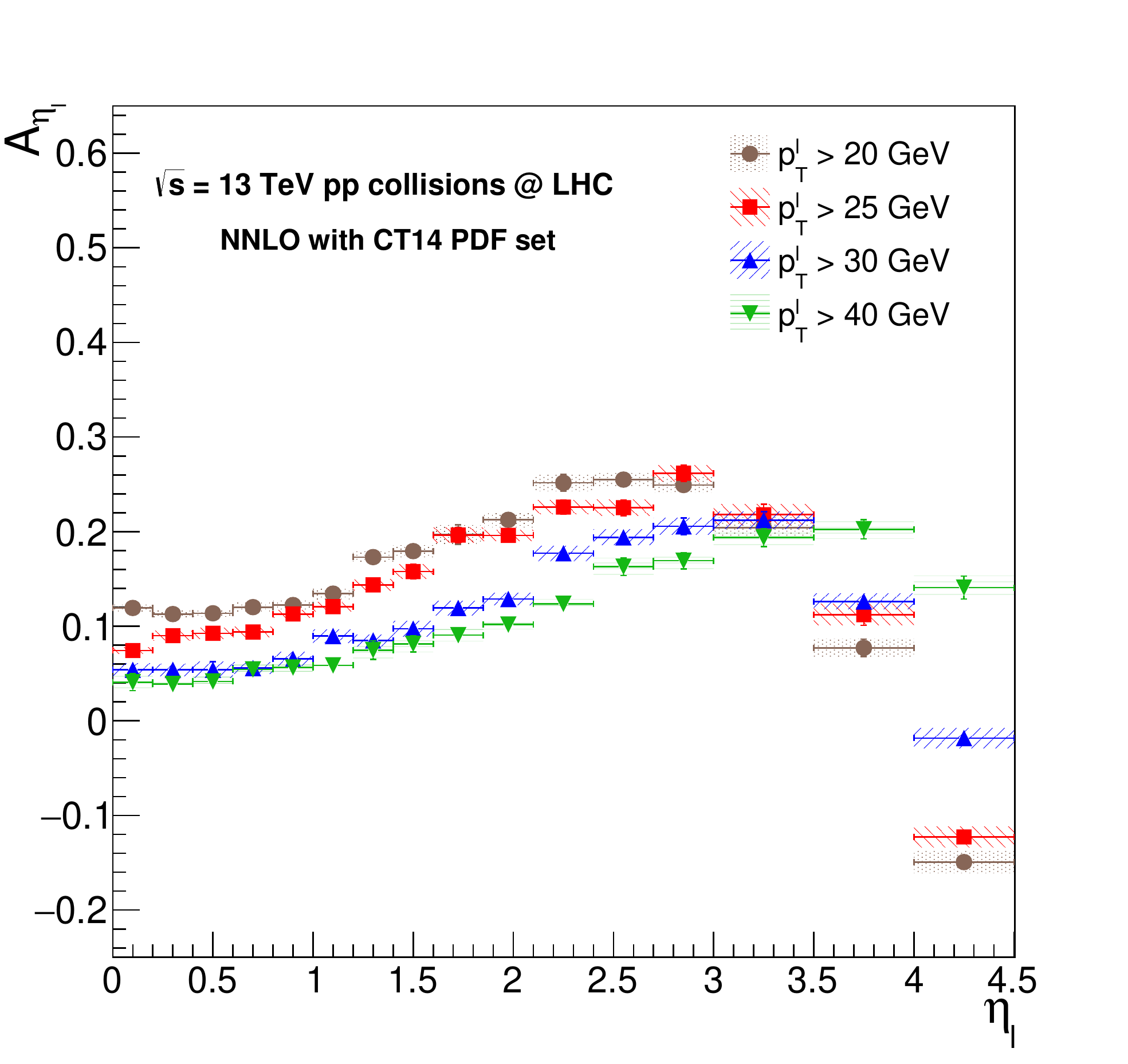}
\caption{The 13 TeV predicted distributions for the $A_{\eta_{l}}$ variable from different low-$p^{l}_{T}$ thresholds $p^{l}_{T} >$ 20, 25, 30, and 40 GeV in bins of the $\eta_{l}$. The NNLO predictions are obtained in both central and forward regions by using the CT14 PDF set. Total theoretical uncertainties are also included for the distributions.}
\label{fig:4}      
\end{figure}

The 13 TeV charge asymmetry predictions from the perturbative QCD calculations of $W^{+}$ and $W^{-}$ boson differential cross sections are reported in this section. The predictions are obtained at NNLO accuracy for the $A_{\eta_{l}}$ variable, where $l$ is either $\mu$ or $e$, by employing $p^{l}_{T} >$ 20, 25, 30, and 40 GeV thresholds in both central and forward phase space regions $0 \leq \eta_{l} \leq 4.5$. Total theoretical uncertainties are estimated using the procedure as described in Sec.~\ref{theo}. The CT14 PDF set at NNLO accuracy is used in the calculations. Bin edges for the $\eta_{l}$ are used identically from the 8 TeV CMS measurement for the central region and are chosen for broader ranges to ensure more stable numerical results in the forward region as (0.00, 0.20), (0.20, 0.40), (0.40, 0.60), (0.60, 0.80), (0.80, 1.00), (1.00, 1.20), (1.20, 1.40), (1.40, 1.60), (1.60, 1.85), (1.85, 2.10), (2.10, 2.40), (2.40, 2.70), (2.70, 3.00), (3.00, 3.50), (3.50, 4.00), and (4.00, 4.50). The predicted $A_{\eta_{l}}$ distributions from different low-$p^{l}_{T}$ thresholds are shown in Fig.~\ref{fig:4}. The predicted $A_{\eta_{l}}$ numerical values corresponding to Fig.~\ref{fig:4} are also given in Table~\ref{tab:1}. The $A_{\eta_{l}}$ distributions increase towards to $\eta_{l}$ bin 3.00--3.50 where they start to turn down for lower values through very forward bins. The $A_{\eta_{l}}$ distribution clearly exhibits dependency on the minimum value of the $p^{l}_{T}$ in both central and forward regions. The $A_{\eta_{l}}$ values decrease in going from a lower $p^{l}_{T}$ threshold to a higher $p^{l}_{T}$ threshold in the central region, while this correlation is reversed in the most forward two bins 3.50--4.50. This is to say lepton charge asymmetry is higher in the central region while it decreases more rapidly in the most forward region when using a lower $p^{l}_{T}$ threshold. 

\begin{figure}
\includegraphics[width=8.8cm]{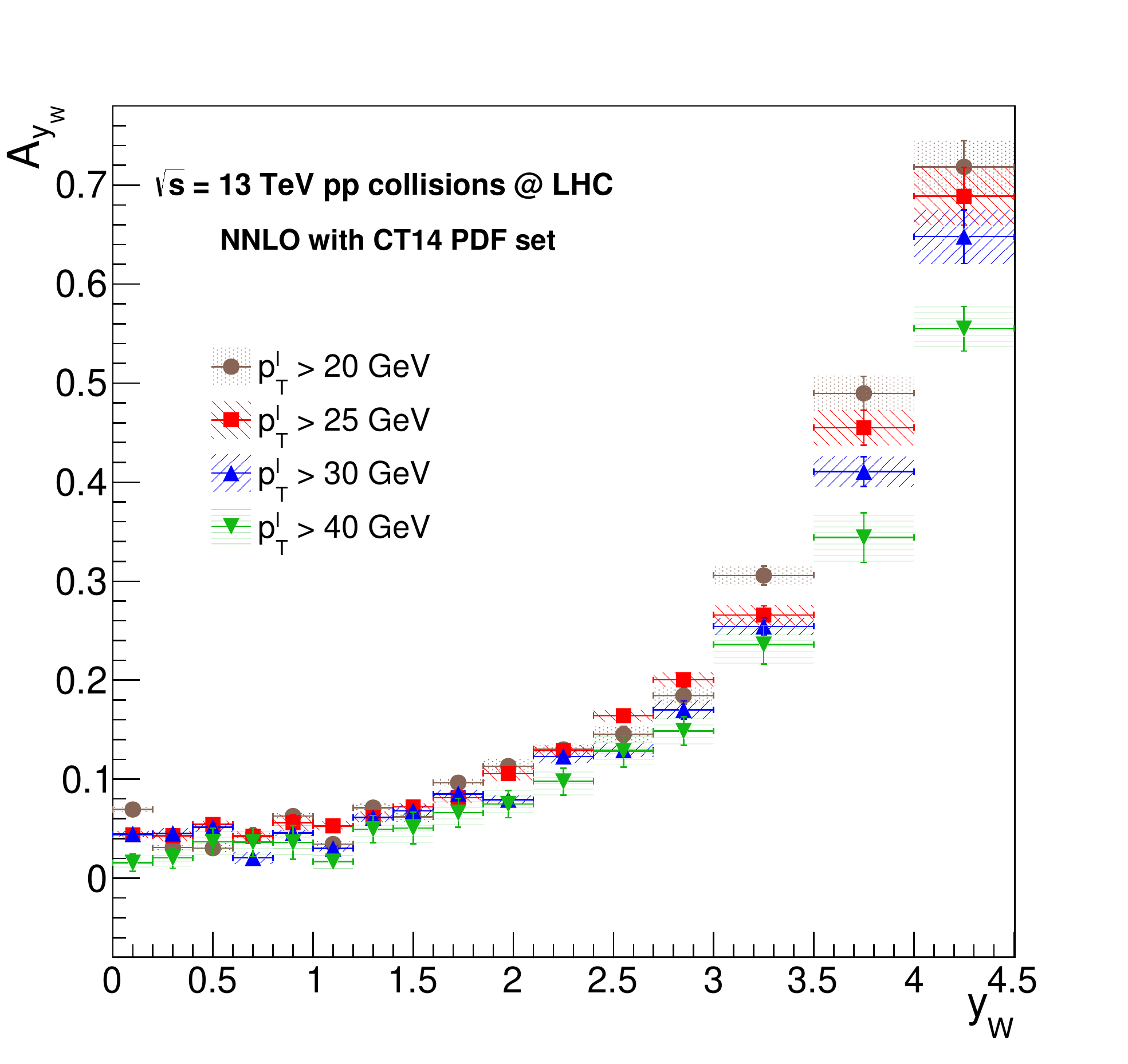}
\caption{The 13 TeV predicted distributions for the $A_{y_{W}}$ variable from different low-$p^{l}_{T}$ thresholds $p^{l}_{T} >$ 20, 25, 30, and 40 GeV in bins of the $y_{W}$. The NNLO predictions are obtained in both central and forward regions by using the CT14 PDF set. Total theoretical uncertainties are also included for the distributions.}
\label{fig:5}      
\end{figure}

The charge asymmetry predictions are also obtained directly for the $A_{y_{W}}$ variable at NNLO accuracy as a function of the $y_{W}$, where $y_{W}$ is calculated from the rapidities of the decay lepton and neutrino. Similarly, fiducial phase space requirements and $y_{W}$ bin edges for the $A_{y_{W}}$ predictions are used from the $A_{\eta_{l}}$ predictions at 13 TeV. The predicted $A_{y_{W}}$ distributions from different low-$p^{l}_{T}$ thresholds are compared as shown in Fig.~\ref{fig:5}. The predicted numerical values from Fig.~\ref{fig:5} are also provided in Table~\ref{tab:2}. The $A_{y_{W}}$ distributions increase consistently towards higher ranges of the $y_{W}$ regardless of the $p^{l}_{T}$ threshold. Contrary to the $A_{\eta_{l}}$ variable, the $A_{y_{W}}$ variable does not discriminate clearly among the predictions from different $p^{l}_{T}$ thresholds in the central region. However the prediction with $p^{l}_{T} >$ 40 GeV tends to be slightly lower than other predictions in the central region. The distributions start to increase more rapidly for a lower threshold in the forward bins, where the distribution with $p^{l}_{T} >$ 40 GeV threshold is predicted to be the most lower one in those bins. The $A_{y_{W}}$ distribution increases the most with lowest threshold $p^{l}_{T} >$ 20 GeV in the forward bins. The correlation between $A_{\eta_{l}}$ and $A_{y_{W}}$ variables become more apparent in the forward region when the distribution shapes approach each other in the presence of increasing values of the $p^{l}_{T}$. Therefore, higher $p^{l}_{T}$ threshold relates $A_{\eta_{l}}$ variable to $A_{y_{W}}$ variable increasingly more in the forward region. The $A_{\eta_{l}}$ distribution with the highest $p^{l}_{T}$ threshold also approaches to the $A_{y_{W}}$ distribution more in the central region. This observation can be supported by the explanation that the average angle between the $W$ boson and the decay lepton decreases when $p^{l}_{T}$ is increased. As a result, the correlation between $A_{\eta_{l}}$ and $A_{y_{W}}$ variables is enhanced. The $A_{\eta_{l}}$ distribution using a higher threshold in the low-$p^{l}_{T}$ region probes the $A_{y_{W}}$ more by allowing a finer dependence on the $x$ values, where a unique set of inputs for the PDF determination can be obtained. Finally, the total theoretical uncertainty of the $A_{y_{W}}$ prediction at NNLO is compared with the total experimental uncertainty of a recent measurement preformed at the LHC~\cite{Sirunyan:2020oum}, in which the W boson asymmetry is reported for the $y_{W}$ at 13 TeV. The theoretical uncertainties, which are in the range $\sim$2--14\% in the presence of the threshold $p^{l}_{T} >$ 25 from Table~\ref{tab:2}, are found to be smaller than or at most comparable to the total experimental uncertainty of the $A_{y_{W}}$ measurement in the $0 \leq y_{W} \leq 2.5$ range. The total experimental uncertainty increases towards higher ranges in the central $y_{W}$ region, and therefore, $A_{y_{W}}$ measurements are anticipated to be challenging in the forward region $2.0 \leq y_{W} \leq 4.5$ in terms of experimental precision to be achieved.                 

\begin{table*}
\caption{The predicted values for the lepton charge asymmetry (in percent) $A_{\eta_{l}}$(\%) at NNLO accuracy by using CT14 PDF sets at 13 TeV. The predictions are reported for different $p^{l}_{T}$ thresholds in bins of the $\eta_{l}$. The predictions include total theoretical uncertainties.}
\label{tab:1}    
\centering
\begin{tabular}{ccccc}
\hline\noalign{\smallskip}
$\eta_{l}$ & $p^{l}_{T} >$ 20 GeV & $p^{l}_{T} >$ 25 GeV & $p^{l}_{T} >$ 30 GeV & $p^{l}_{T} >$ 40 GeV \\
\noalign{\smallskip}\hline\noalign{\smallskip}
0.00--0.20  & 11.93$\pm$0.4     & 07.43$\pm$0.4        & 05.39$\pm$0.7    & 04.09$\pm$0.9  \\
0.20--0.40  & 11.28$\pm$0.4     & 09.00$\pm$0.5        & 05.39$\pm$0.6    & 03.89$\pm$0.4 \\
0.40--0.60  & 11.37$\pm$0.3     & 09.25$\pm$0.6        & 05.40$\pm$0.8    & 04.17$\pm$0.6 \\
0.60--0.80  & 12.02$\pm$0.5     & 09.39$\pm$0.5        & 05.74$\pm$0.6    & 05.44$\pm$0.5 \\
0.80--1.00  & 12.24$\pm$0.5     & 11.30$\pm$0.4        & 06.55$\pm$0.7    & 05.64$\pm$0.7 \\
1.00--1.20  & 13.45$\pm$0.6     & 12.07$\pm$0.5        & 08.96$\pm$0.7     & 05.87$\pm$0.5 \\
1.20--1.40  & 17.31$\pm$0.6     & 14.37$\pm$0.6        & 08.50$\pm$0.6     & 07.48$\pm$0.9 \\
1.40--1.60  & 17.94$\pm$0.6     & 15.78$\pm$0.7        & 09.74$\pm$0.8     & 08.14$\pm$0.9  \\
1.60--1.85  & 19.70$\pm$1.0     & 19.63$\pm$0.8        & 11.93$\pm$0.7     & 09.05$\pm$0.7  \\
1.85--2.10  & 21.25$\pm$0.8     & 19.61$\pm$0.6        & 12.87$\pm$0.6     & 10.17$\pm$0.5 \\
2.10--2.40  & 25.16$\pm$0.9     & 22.60$\pm$0.7        & 17.71$\pm$0.8     & 12.37$\pm$0.6  \\
2.40--2.70  & 25.50$\pm$0.7     & 22.53$\pm$0.8        & 19.40$\pm$0.8     & 16.29$\pm$0.9 \\
2.70--3.00  & 24.93$\pm$0.8     & 26.17$\pm$0.8        & 20.57$\pm$0.8     & 16.92$\pm$0.8 \\
3.00--3.50  & 20.42$\pm$1.0     & 21.79$\pm$1.1        & 21.21$\pm$0.9     & 19.37$\pm$0.9 \\
3.50--4.00  & 07.72$\pm$0.9     & 11.22$\pm$1.1        & 12.62$\pm$0.8     & 20.25$\pm$1.0 \\
4.00--4.50  & -14.93$\pm$1.1     & -12.27$\pm$1.1        & -01.82$\pm$1.0     & 14.09$\pm$1.2  \\
\noalign{\smallskip}\hline
\end{tabular}
\end{table*}

\begin{table*}
\caption{The predicted values for the $W$ boson charge asymmetry (in percent) $A_{y_{W}}$(\%) at NNLO accuracy by using CT14 PDF sets at 13 TeV. The predictions are reported for different $p^{l}_{T}$ thresholds in bins of the $y_{W}$. The predictions include total theoretical uncertainties.}
\label{tab:2}    
\centering
\begin{tabular}{ccccc}
\hline\noalign{\smallskip}
$y_{W}$ & $p^{l}_{T} >$ 20 GeV & $p^{l}_{T} >$ 25 GeV & $p^{l}_{T} >$ 30 GeV & $p^{l}_{T} >$ 40 GeV \\
\noalign{\smallskip}\hline\noalign{\smallskip}
0.00--0.20  & 06.93$\pm$0.3     & 04.38$\pm$0.4        & 04.46$\pm$0.3    & 01.55$\pm$0.9  \\
0.20--0.40  & 03.08$\pm$0.3     & 04.28$\pm$0.5        & 04.52$\pm$0.5    & 02.04$\pm$1.0 \\
0.40--0.60  & 03.03$\pm$0.5     & 05.42$\pm$0.4        & 05.13$\pm$0.4    & 03.66$\pm$1.3 \\
0.60--0.80  & 04.30$\pm$0.2     & 04.23$\pm$0.5        & 02.05$\pm$0.6    & 03.64$\pm$1.5 \\
0.80--1.00  & 06.27$\pm$0.3     & 05.60$\pm$0.8        & 04.57$\pm$0.3    & 03.61$\pm$1.7 \\
1.00--1.20  & 03.44$\pm$0.4     & 05.27$\pm$0.4        & 03.00$\pm$0.3     & 01.66$\pm$0.8 \\
1.20--1.40  & 07.12$\pm$0.5     & 06.14$\pm$0.5        & 06.12$\pm$0.4     & 04.95$\pm$1.4 \\
1.40--1.60  & 06.19$\pm$0.5     & 07.20$\pm$0.4        & 06.80$\pm$0.5     & 05.07$\pm$1.5  \\
1.60--1.85  & 09.64$\pm$0.4     & 08.13$\pm$0.5        & 08.51$\pm$0.4     & 06.61$\pm$1.5  \\
1.85--2.10  & 11.32$\pm$0.7     & 10.56$\pm$0.7        & 07.93$\pm$0.4     & 07.47$\pm$1.4 \\
2.10--2.40  & 13.00$\pm$0.5     & 12.90$\pm$0.5        & 12.30$\pm$0.6     & 09.75$\pm$1.4  \\
2.40--2.70  & 14.52$\pm$0.8     & 16.40$\pm$0.5        & 12.92$\pm$0.7     & 12.83$\pm$1.6 \\
2.70--3.00  & 18.42$\pm$0.8     & 20.03$\pm$0.7        & 16.99$\pm$0.9     & 14.85$\pm$1.5 \\
3.00--3.50  & 30.57$\pm$0.9     & 26.57$\pm$0.9        & 25.42$\pm$0.8     & 23.59$\pm$1.9 \\
3.50--4.00  & 48.97$\pm$1.7     & 45.50$\pm$1.8        & 41.06$\pm$1.5     & 34.40$\pm$2.5 \\
4.00--4.50  & 71.84$\pm$2.6     & 68.87$\pm$2.9        & 64.80$\pm$2.7     & 55.50$\pm$2.3  \\
\noalign{\smallskip}\hline
\end{tabular}
\end{table*}

\section{The predictions in the $p^{W}_{T}$ bins at 13 TeV}
\label{13tevptw}
\begin{figure}
\includegraphics[width=8.8cm]{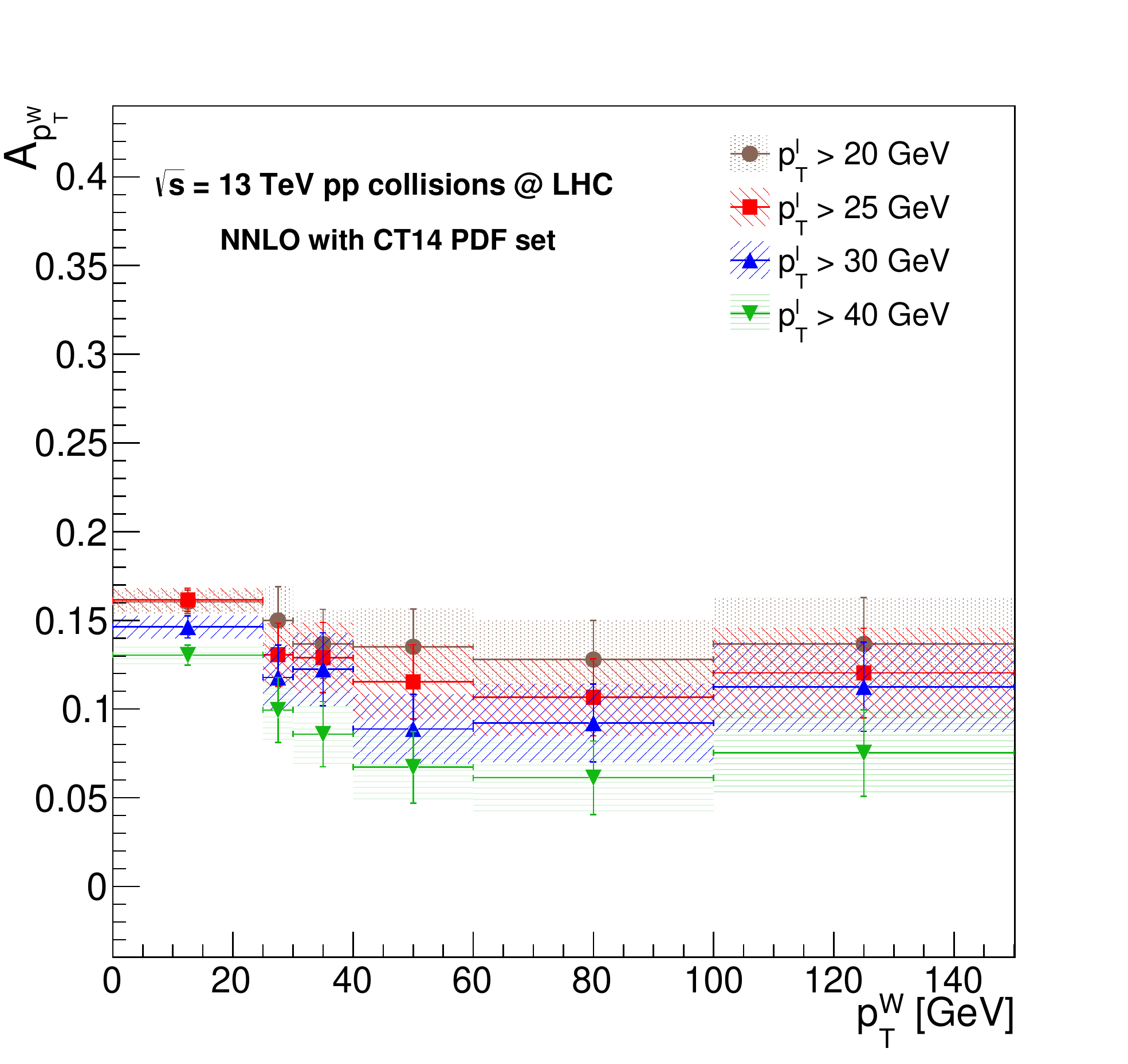}

\caption{The 13 TeV predicted distributions for the $A_{p^{W}_{T}}$ variable based on different low-$p^{l}_{T}$ thresholds $p^{l}_{T} >$ 20, 25, 30, and 40 GeV in bins of the $p^{W}_{T}$. The predictions at NNLO accuracy are obtained in both central and forward regions by using the CT14 PDF set. Total theoretical uncertainties are also included for the distributions.}
\label{fig:6}      
\end{figure}

The charge asymmetry predictions are provided in the previous section for the variables that are defined in terms of the $\eta_{l}$ and $y_{W}$. The charge asymmetry can also be predicted in terms of the $p^{W}_{T}$ by using an analogous definition with regards to Eqns.~\ref{eqn:1} and~\ref{eqn:2} as   

\begin{equation}
\label{eqn:4}   
A_{p^{W}_{T}}=\frac{d\sigma(W^{+}\rightarrow l^{+}\nu)/dp^{W}_{T}-d\sigma(W^{-}\rightarrow l^{-}\bar{\nu})/dp^{W}_{T}}{d\sigma(W^{+}\rightarrow l^{+}\nu)/dp^{W}_{T}+d\sigma(W^{-}\rightarrow l^{-}\bar{\nu})/dp^{W}_{T}}
\end{equation}   
to acquire more insight into the $W$ boson production asymmetry in the presence of different $p^{l}_{T}$ thresholds. The $A_{p^{W}_{T}}$ variable can reveal additional information for the impact of using various $p^{l}_{T}$ thresholds on the predicted asymmetry in bins of the $p^{W}_{T}$. The $A_{p^{W}_{T}}$ predictions at 13 TeV are first obtained at NNLO to test dependency to low-$p^{l}_{T}$ thresholds of $p^{l}_{T} >$ 20, 25, 30, and 40 GeV. The predictions are obtained for the region $0 \leq \eta_{l} \leq 4.5$ in the $p^{W}_{T}$ range 0--150 GeV as shown in Fig.~\ref{fig:6}. The predicted numerical values at NNLO from Fig.~\ref{fig:6} are also provided in Table~\ref{tab:3}. The $A_{p^{W}_{T}}$ distributions consistently decrease towards higher bin ranges of the $p^{W}_{T}$ at all $p^{l}_{T}$ thresholds except for the highest bin, however the distribution is more flat at the lowest threshold of 20 GeV. The $A_{p^{W}_{T}}$ distribution obviously decreases in going from the prediction using a lower threshold to the prediction using a higher threshold in $p^{l}_{T}$ throughout the entire $p^{W}_{T}$ region. Therefore, the increasing $p^{l}_{T}$ threshold has an impact to yield lower $A_{p^{W}_{T}}$ values in both central and forward regions. These results already emphasize that the $p^{W}_{T}$ and $p^{l}_{T}$ are closely correlated kinematical variables. 

\begin{figure}
\includegraphics[width=8.8cm]{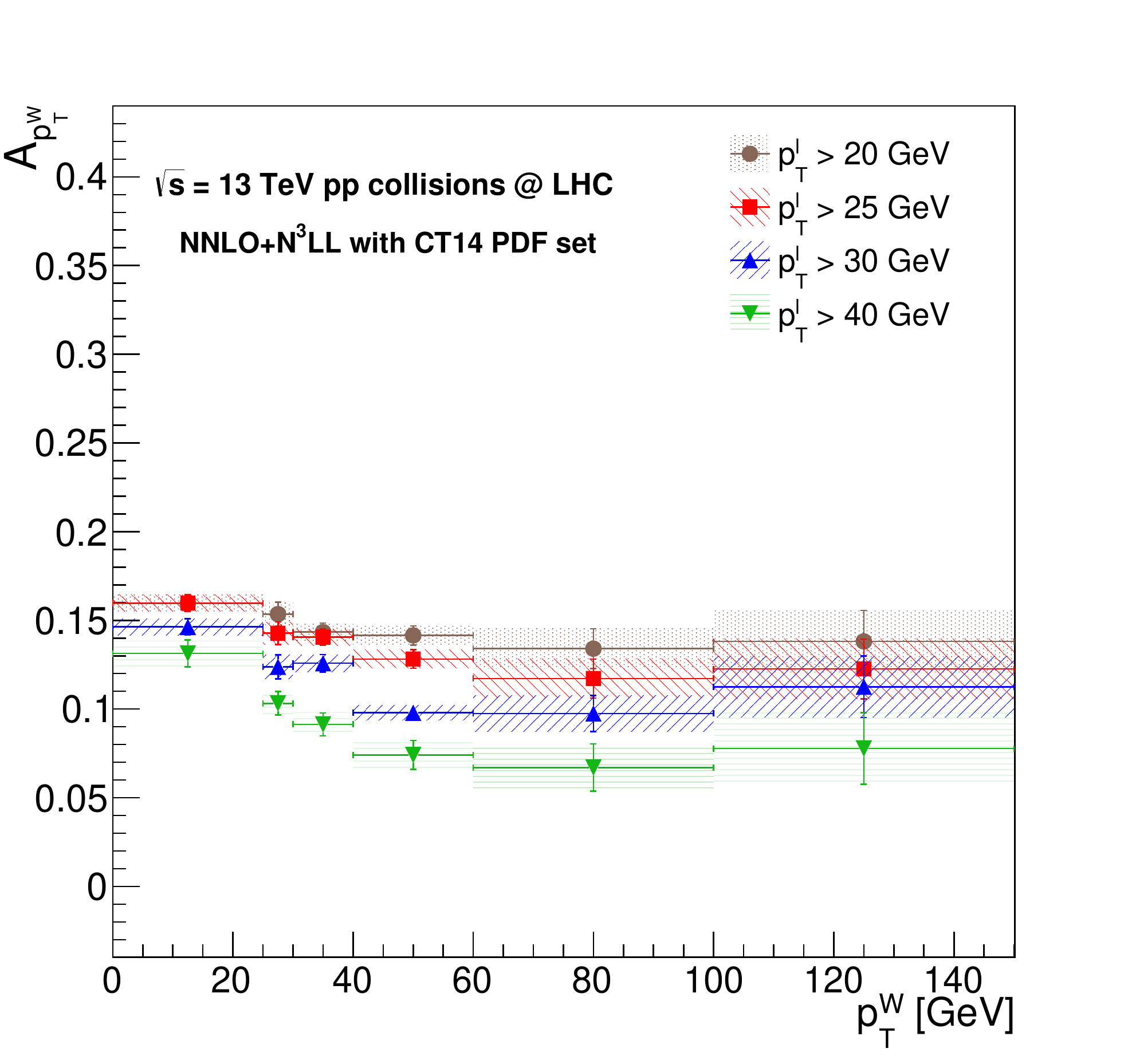}

\caption{The 13 TeV predicted distributions for the $A_{p^{W}_{T}}$ variable based on different low-$p^{l}_{T}$ thresholds $p^{l}_{T} >$ 20, 25, 30, and 40 GeV in bins of the $p^{W}_{T}$. The merged predictions at NNLO+N$^{3}$LL accuracy are obtained in both central and forward regions by using the CT14 PDF set. Total theoretical uncertainties are also included for the distributions.}
\label{fig:7}      
\end{figure}

Next the $A_{p^{W}_{T}}$ distributions at 13 TeV are obtained by means of the matched calculation of the resummation with the fixed-order NNLO to have more accurate predictions. The $p^{W}_{T}$ distribution is affected by soft and collinear gluon radiation at low values, where the fixed-order calculations are unable to sufficiently account for it. Thereby, resummation of large logarithmic corrections are also needed to model $p^{W}_{T}$ more accurately at low values. The differential cross sections of W$^{+}$ and $W^{-}$ bosons as a function of the $p^{W}_{T}$ are predicted by means of the matched prediction at NNLO+N$^{3}$LL accuracy to achieve more reliable $A_{p^{W}_{T}}$ predictions. The $A_{p^{W}_{T}}$ distributions that are predicted at NNLO+N$^{3}$LL accuracy by employing different $p^{l}_{T}$ thresholds are shown in Fig.~\ref{fig:7}. In Table~\ref{tab:4}, the predicted numerical values from Fig.~\ref{fig:7} are also supplied. Similarly, the predicted values decrease continuously with increasing $p^{l}_{T}$ threshold apart from the highest bin, as being also tested at a higher accuracy of NNLO+N$^{3}$LL. The highest threshold yields the lowest $A_{p^{W}_{T}}$ values through the entire $p^{W}_{T}$ ranges. Therefore, it is shown that the $A_{p^{W}_{T}}$ predictions also depend on the $p_{T}$ threshold applied on the decay lepton which can be attributed to the V--A structure of the $W$ boson couplings to fermions. 

To this end, experimental uncertainty of normalized differential cross section as a function of the $p^{W}_{T}$ is checked from the CMS measurement~\cite{Khachatryan:2016nbe} to anticipate a comparison between experimental uncertainty that can be achieved and the theoretical uncertainty estimated at NNLO(+N$^{3}$LL) accuracy for the $A_{p^{W}_{T}}$ at 13 TeV. Experimental uncertainty including both systematic and statistical components is $\sim$1.2--6.0\%, whereas the total theoretical uncertainty ranges in 2.1--9.0\% (1.5--6.0\%) in the NNLO (NNLO+N$^{3}$LL) predictions for differential cross section of the $p^{W}_{T}$ in the range 0-150 GeV. Therefore, experimental uncertainty is anticipated to be lower than (comparable to) the theoretical uncertainty in the NNLO (NNLO+N$^{3}$LL) predictions of the $A_{p^{W}_{T}}$ at 13 TeV.         

\begin{table*}
\caption{The predicted values for the charge asymmetry (in percent) $A_{p^{W}_{T}}$(\%) at NNLO accuracy by using CT14 PDF set at 13 TeV. The predictions are reported for different $p^{l}_{T}$ thresholds in bins of the $p^{W}_{T}$. The predictions include total theoretical uncertainties.}
\label{tab:3}    
\centering
\begin{tabular}{ccccc}
\hline\noalign{\smallskip}
$p^{W}_{T}$ & $p^{l}_{T} >$ 20 GeV & $p^{l}_{T} >$ 25 GeV & $p^{l}_{T} >$ 30 GeV & $p^{l}_{T} >$ 40 GeV \\
\noalign{\smallskip}\hline\noalign{\smallskip}
0--25  & 16.04$\pm$0.7     & 16.15$\pm$0.6        & 14.63$\pm$0.6    & 13.04$\pm$0.6  \\
25--30  & 15.00$\pm$1.9     & 13.06$\pm$1.8        & 11.78$\pm$1.8    & 09.94$\pm$1.8 \\
30--40  & 13.67$\pm$1.9     & 12.90$\pm$1.9        & 12.24$\pm$2.0    & 08.58$\pm$1.8 \\
40--60  & 13.50$\pm$2.1     & 11.53$\pm$2.1        & 08.88$\pm$1.9    & 06.72$\pm$2.0 \\
60--100  & 12.79$\pm$2.2     & 10.66$\pm$2.2        & 09.20$\pm$2.2    & 06.12$\pm$2.0 \\
100--150  & 13.67$\pm$2.6     & 12.03$\pm$2.5        & 11.25$\pm$2.5     & 07.52$\pm$2.4 \\
\noalign{\smallskip}\hline
\end{tabular}
\end{table*}

\begin{table*}
\caption{The predicted values for the charge asymmetry (in percent) $A_{p^{W}_{T}}$(\%) at NNLO+N$^{3}$LL accuracy by using CT14 PDF set at 13 TeV. The predictions are reported for different $p^{l}_{T}$ thresholds in bins of the $p^{W}_{T}$. The predictions include total theoretical uncertainties.}
\label{tab:4}    
\centering
\begin{tabular}{ccccc}
\hline\noalign{\smallskip}
$p^{W}_{T}$ & $p^{l}_{T} >$ 20 GeV & $p^{l}_{T} >$ 25 GeV & $p^{l}_{T} >$ 30 GeV & $p^{l}_{T} >$ 40 GeV \\
\noalign{\smallskip}\hline\noalign{\smallskip}
0--25  & 15.97$\pm$0.5     & 15.97$\pm$0.5        & 14.62$\pm$0.5    & 13.13$\pm$0.8  \\
25--30  & 15.35$\pm$0.7     & 14.27$\pm$0.6        & 12.38$\pm$0.7    & 10.32$\pm$0.7 \\
30--40  & 14.34$\pm$0.5     & 14.05$\pm$0.5        & 12.58$\pm$0.5    & 09.13$\pm$0.6 \\
40--60  & 14.15$\pm$0.5     & 12.82$\pm$0.5        & 09.79$\pm$0.4    & 07.40$\pm$0.8 \\
60--100  & 13.40$\pm$1.1     & 11.72$\pm$1.1        & 09.75$\pm$1.0    & 06.70$\pm$1.3 \\
100--150  & 13.81$\pm$1.7     & 12.26$\pm$1.7        & 11.25$\pm$1.7     & 07.77$\pm$2.0 \\
\noalign{\smallskip}\hline
\end{tabular}
\end{table*}

\section{Conclusion}
\label{conclusion}
In this paper a dedicated study of the theoretical predictions for the $W$ boson charge asymmetries in $pp$ collisions is presented. The predictions are obtained with the inclusion of NNLO corrections in perturbative QCD for both the central and forward phase space regions $0 \leq \eta_{l} \leq 4.5$. This phase space region for the charge asymmetries allows to probe the relative $u$ and $d$ quark distributions in the proton at very small and large $x$ values. The predictions that are based on various PDF models are compared with the CMS and LHCb data for the lepton charge asymmetry $A_{\eta_{l}}$ in the muon decay mode at 8 TeV. The $A_{\eta_{l}}$(\%) distributions are generally observed to be in good agreement with the 8 TeV data within the quoted uncertainties. The distributions using the CT14 PDF set are shown to reproduce the data slightly better over the other PDF sets that are being tested. These comparisons enabled the justification of the predictions and encouraged the extension of the study to 13 TeV which is the current center-of-mass energy of $pp$ collisions at the LHC.         

The predicted distributions are presented for the charge asymmetries $A_{\eta_{l}}$ and $A_{y_{W}}$ at NNLO accuracy at 13 TeV. Various increasing low-$p^{l}_{T}$ thresholds $p^{l}_{T} >$ 20, 25, 30, and 40 GeV in the region $0 \leq \eta_{l} \leq 4.5$ are used to assess the impact on the predicted asymmetries. The $A_{\eta_{l}}$ distributions from different $p^{l}_{T}$ thresholds increase in the central region and decrease in the forward region. The $A_{\eta_{l}}$ distribution exhibits clear dependency to $p^{l}_{T}$, where it increases more in correlation with decreasing threshold in the central region up to the $\eta_{l}$ bin 3.0--3.5. After this bin, the $A_{\eta_{l}}$ distribution decreases more with lower threshold in the very forward range 3.5--4.5. The predicted $A_{\eta_{l}}$ is observed to be the lowest (highest) in the central (forward) region with the highest threshold $p^{l}_{T} >$ 40 GeV that is being tested. Furthermore, the charge asymmetry distributions that are obtained directly with the $A_{y_{W}}$ are observed to increase continuously towards higher ranges of the $y_{W}$. The $A_{y_{W}}$ is predicted to be larger at higher $y_{W}$ bins which can be attributed to increasing ratio of the $u$ and $d$ quark distribution functions while probing the valence quarks more in those higher bins. Contrary to the $A_{\eta_{l}}$, the $A_{y_{W}}$ distribution does not strongly discriminate among different thresholds in most of the central ranges. The $A_{y_{W}}$ distribution increases more in correlation with lower threshold in the forward $y_{W}$ bins, where it is lower with the $p^{l}_{T} >$ 40 GeV threshold. In the $A_{\eta_{l}}$ and $A_{y_{W}}$ predictions, it has been clearly shown that $A_{\eta_{l}}$ distribution gets closer to the $A_{y_{W}}$ distribution in the presence of the highest threshold $p^{l}_{T} >$ 40 GeV in both the central and forward regions. This observation is in support of the point that the average angle between the $W$ boson and the decay lepton is decreased with a higher $p^{l}_{T}$ requirement, and as a result, the correlation between the $A_{\eta_{l}}$ and $A_{y_{W}}$ is enhanced accordingly.              

The 13 TeV distributions are also reported for the charge asymmetry in bins of the $p^{W}_{T}$, $A_{p^{W}_{T}}$, at both NNLO and NNLO+N$^{3}$LL, where the accuracy is remarkably improved in the matched predictions. The predicted distributions decrease continuously towards the higher $p^{W}_{T}$ ranges apart from the very last bin at 100--150 GeV. The $A_{p^{W}_{T}}$ distribution is observed to exhibit clear dependency to $p^{l}_{T}$ threshold as anticipated. It has been shown that the distribution remains more flat at the lowest threshold of 20 GeV while it decreases more in correlation with the increasing threshold. Based on the results presented, it was shown that the $A_{p^{W}_{T}}$ can be used as an alternate probe for the $W$ boson charge asymmetries.       

Finally, the study shows the potential impact of the $p^{l}_{T}$ dependence in the $W$ boson leptonic decay in terms of the charge asymmetries. The predicted results can further be used for improving the existing constraints on the ratio of $u$ and $d$ quark distribution functions in the range $10^{-4} < x < 1$, investigating into differences among PDF models, contribution in general to accurate PDF determinations.  


\section*{Conflict of interest}
The author declares that he has no conflict of interest.


\end{document}